\documentclass[12pt]{article}
\usepackage[dvips]{graphics}
\usepackage{epsfig}

\catcode`@=11
\def\@citex[#1]#2{\if@filesw\immediate\write\@auxout{\string\citation{#2}}\fi
  \def\@citea{}\@cite{\@for\@citeb:=#2\do
    {\@citea\def\@citea{,\penalty\@m}\@ifundefined
      {b@\@citeb}{{\bf ?}\@warning
       {Citation `\@citeb' on page \thepage \space undefined}}%
\hbox{\csname b@\@citeb\endcsname}}}{#1}}

\def\citer{\@ifnextchar
[{\@tempswatrue\@citexr}{\@tempswafalse\@citexr[]}}
\def\@citexr[#1]#2{\if@filesw\immediate\write\@auxout{\string\citation{#2}}\fi
  \def\@citea{}\@cite{\@for\@citeb:=#2\do
    {\@citea\def\@citea{--\penalty\@m}\@ifundefined
       {b@\@citeb}{{\bf ?}\@warning
       {Citation `\@citeb' on page \thepage \space undefined}}%
\hbox{\csname b@\@citeb\endcsname}}}{#1}}
\catcode`@=12

\newcommand{\gsim}{\buildrel > \over {_\sim}}

\newcommand{\be}{\begin{equation}}
\newcommand{\ee}{\end{equation}}
\newcommand{\bea}{\begin{eqnarray}}
\newcommand{\eea}{\end{eqnarray}}
\newcommand{\ba}{\begin{array}}
\newcommand{\ea}{\end{array}}

\newcommand{\etal}{{\em et al.}}
\newcommand{\ibid}{{\em ibid.}}

\newcommand{\msbar}{$\overline{\rm MS}$}

\def\alphat{\hat\alpha}

\begin{document}

\hfill
\begin{tabular}{r}
{\normalsize FT--2004--03} \\
{\normalsize Caltech MAP--300}
\end{tabular}

\vspace{18pt}
\centerline{\Large\bf The Weak Mixing Angle at Low Energies}
\vspace{18pt}
\centerline{\sc Jens Erler$^{\;1}$ and Michael J. Ramsey-Musolf$^{\;2}$}
\vspace{6pt}
\centerline{\it $^1$Instituto de F\'\i sica, Universidad Nacional Aut\'onoma
                de M\'exico, 01000 M\'exico D.F., M\'exico}
\vspace{6pt}
\centerline{\it $^2$Kellogg Radiation Laboratory, California Institute of
                Technology, Pasadena, CA 91125, USA}

\vspace{18pt}

\begin{abstract}
We determine the weak mixing angle in the $\overline{\rm MS}$-scheme,  $\sin^2\hat\theta_W(\mu)$, at energy scales $\mu$ relevant for present and  future low energy electroweak measurements.  We relate the renormalization group evolution of  $\sin^2\hat\theta_W(\mu)$ to the corresponding evolution of ${\hat\alpha}(\mu)$ and include higher-order terms in $\alpha_s$ and $\alpha$ that had not been treated in previous analyses. We also up-date the analysis of non-perturbative, hadronic contributions and argue that the associated uncertainty is  small compared to anticipated experimental errors. The resulting value of the low-energy  $\overline{\rm MS}$ weak mixing angle is $\sin^2\hat\theta_W(0) = 0.23867 \pm 0.00016$. 
\end{abstract}

\vspace{0pt}

\section{Introduction}
\label{sec:intro}
The weak mixing angle is one of the fundamental parameters of the electroweak 
Standard Model (SM). It can be defined through the relation,
\be
   \sin^2\theta_W = {{g^\prime}^2\over g^2 + {g^\prime}^2},
\label{sin2t}
\ee
where $g$ and $g^\prime$ are the $SU(2)_L$ and $U(1)_Y$ gauge couplings,
respectively.  Its value is not predicted and needs to be extracted from parity
violating neutral current experiments, where by far the most precise results 
were obtained at the $Z$ factories LEP~1 and SLC. Electroweak symmetry breaking
provides masses for the $W$ and $Z$ bosons proportional to their gauge 
interactions.  Therefore, one has the additional relation,
\be 
   \sin^2\theta_W = 1 - {M_W^2\over M_Z^2}, 
\label{onshell}
\ee
and the gauge boson mass ratio provides independent precise information on 
$\sin^2\theta_W$.  Extracting the fine structure constant,
\be
   \alpha = {e^2\over 4\pi} = {g^2 \sin^2\theta_W\over 4\pi},
\label{alpha}
\ee
from the quantum Hall effect or the anomalous magnetic moment of the electron,
then fixes both gauge couplings. Eqs.~(\ref{sin2t}) and (\ref{alpha}) are valid
at the tree level and modified by radiative corrections. As a result, 
the precise numerical value of $\sin^2\theta_W$ depends on the renormalization 
scheme and scale chosen. For example, the on-shell renormalization scheme 
promotes Eq.~(\ref{onshell}) to a definition of 
$\sin^2\theta_W \equiv \sin^2\theta_W^{\rm \; on-shell}$ to all orders in 
perturbation theory. This definition has the advantage of being directly 
related to a physical observable --- but only to one-loop order. Since 
the gauge bosons are unstable particles their masses become ambiguous starting 
at two-loop precision.  

As an alternative, one can  define flavor-dependent ``effective'' 
mixing angles appearing in the $Z$ vector coupling\footnote{We normalize $v_f$
without an additional factor inversely proportional to 
$\sin\theta_W\cos\theta_W$. The scale in this factor needs to be chosen of 
electroweak size, $\mu = M_Z$, even for low energy processes. This is 
automatically achieved by normalizing neutral-current amplitudes using 
$\rho\; G_F$, where $G_F$ is the Fermi constant and $\rho$ the low energy 
neutral-current $\rho$ parameter~\cite{Veltman:1977kh} (which is free of 
fermionic mass singularities). Thus, this factor does not affect our 
discussion.},
\be
\label{eq:vf}
   v_f = T_f - 2 Q_f \sin^2\theta_f^{\rm \; eff},
\ee
where $Q_f$ and $T_f$ are the fermion charge and third component of isospin,
respectively. Gauge boson self-energy and $Zf{\bar f}$-vertex corrections are 
absorbed into scheme dependent form factors, $\rho_f$ and $\kappa_f$ that are equal to unity at tree-level. The $\rho_f$ are corrections to the overall coupling strengths and 
the $\kappa_f$ are defined by,
\be
\label{eq:sin2eff}
   \sin^2\theta_f^{\rm \; eff} \equiv \kappa_f \sin^2\theta_W^{\rm \; on-shell}
   \equiv \hat{\kappa}_f \sin^2\hat\theta_W (M_Z),
\ee 
where the caret marks quantities in the modified minimal subtraction 
($\overline{\rm MS}$) scheme~\citer{Bardeen:1978yd,Marciano:1980be}.
This effective $\sin^2\theta_f^{\rm \; eff}$ is a useful definition as long as 
electroweak box contributions can be neglected; since these do not resonate 
this condition is clearly satisfied at LEP~1 and SLC. However, with the greater 
precision that could be achieved with the GigaZ option of TESLA, such boxes 
could become non-negligible. Thus, it is not easy to construct 
a definition that can be equated to a physical observable to all orders. 
Neither is this of practical relevance: as long as it is well-defined,
$\sin^2\theta_W$ can be looked upon as a mere bookkeeping device and means 
to compare various experimental results. 

What is of practical importance, however, is the {\em numerical value\/} of 
the mixing angle used in computing a given observable in a specified renormalization scheme. 
Generally speaking, one expects a one-loop radiatively corrected result to be valid up to small 
corrections of ${\cal O}({\alpha^2\over\pi^2\sin^4\theta_W}) \sim 10^{-4}$. On 
the other hand, for leptons we have,
\be
   (\kappa_\ell - 1)^2 \approx 1.5 \times 10^{-3},
\label{kappae}
\ee
which is not much smaller than a typical one-loop contribution. The reason is 
that $\kappa_\ell$ contains top-quark mass enhancements factors, $m_t^2/M_W^2$, that spoil the expected behavior of the perturbation series. Fortunately, such enhancement factors can be avoided by a judicious choice of definition of the weak mixing angle (or renormalization scheme), rendering the truncation error small. The $\overline{\rm MS}$-definition considered in this article has this 
property, except that small $\ln m_t^2/M_W^2$ corrections cannot be decoupled 
simultaneously from all observables~\cite{Marciano:1990dp,Fanchiotti:1992tu}.

For processes off the $Z$-pole, enhancement factors of similar magnitude as those entering Eq.~(\ref{kappae}) can arise from large logarithms $\ln M_Z^2/m_f^2$, where $m_f$ is some fermion
mass. These can occur even within a specifically chosen renormalization scheme.
Typically, such logarithms are artifacts of using a value for $\sin^2\theta_W$
obtained at the $Z$ scale (where as noted above, it is measured very precisely)
in theoretical expressions for very low (or high) energy observables. In some 
renormalization schemes, the weak mixing angle depends explicitly on 
a renormalization scale parameter, which could be the t'Hooft scale, $\mu$,
appearing in dimensional regularization, or the momentum transfer $q^2$. For 
others there is no explicit scale parameter, and the scale dependence is of 
indirect nature. In either case, the aforementioned logarithms are a potential
hazard and should be avoided or re-summed if possible. 

The main goal of this paper is to present an analysis of the weak mixing angle 
in the $\overline{\rm MS}$-scheme, $\sin^2\hat\theta_W(0)$, relevant for 
observables measured at (almost) vanishing momentum transfer. These observables include the nuclear weak charge obtained in the well-known cesium atomic parity violation
measurement performed by the Boulder group~\cite{Wood:1997zq}; the parity violating M\o ller asymmetry at SLAC~\cite{Anthony:2003ub}; and the up-coming measurement of the weak charge of
the proton at Jefferson Lab~\cite{Armstrong:2003gp}. Precision measurements of 
the parity violating deep inelastic $eD$ asymmetry have also been discussed as 
future possibilities for Jefferson Lab, as has a more precise measurement of the M\o ller asymmetry. 
The value of $\sin^2\hat\theta_W(0)$ is particularly relevant for the interpretation of the parity violation experiments, since the vector coupling of the $Z$ n Eq.~(\ref{eq:vf}) to electrons and protons is proportional to $1-4\sin^2\theta_W(q^2)^{\rm eff}_f\sim 0.1$ and is, therefore, highly sensitive to the value of the effective weak mixing angle. Indeed, as noted in Refs.~\cite{Czarnecki:1995fw,Czarnecki:2000ic}, one-loop contributions to $\sin^2\theta_W(0)^{\rm eff}_f$ reduce the magnitude of the M\o ller asymmetry by roughly 40\% from its tree-level value, an effect generated by the large logarithms discussed above.
Moreover,  the presence of these large logarithms in $\sin^2\theta_W(0)^{\rm eff}_f$ is universal to all low energy neutral current observables, though their net effect may be masked by other enhanced radiative corrections\footnote{ For example, the left-right asymmetry for 
parity-violating  elastic $ep$ scattering also contains large logarithms 
associated with the fermion anapole moment as well as non-logarithmic but large $WW$ box graphs~\cite{Musolf:1990ts,Musolf:1990sa,Marciano:1980pb}.}. Consequently, one would like to sum these universally enhanced contributions to all orders. Here, we do so using the renormalization  group evolution  for $\sin^2\hat\theta_W(\mu)$ in the $\overline{\rm MS}$-scheme.

The $\overline{\rm MS}$-definition of the weak mixing angle is, of course, not 
unique, and one may choose an alternate scheme in which to compute radiative 
corrections to electroweak observables. Nevertheless, 
the $\overline{\rm MS}$-scheme offers several advantages that motivate our 
adoption of it here. In particular, the $\overline{\rm MS}$-definition of 
$\sin^2\hat\theta_W(\mu)$ follows closely the coupling-based definition in 
Eq.~(\ref{sin2t}) with a well-defined subtraction of singular terms arising in 
dimensional regularization, giving rise to expressions with a logarithmic 
$\mu$-dependence. This dependence is governed by a renormalization group 
equation (RGE), and choosing $\mu$ equal to the momentum transfer of 
the process under consideration will in general avoid spurious 
logarithms\footnote{The presence of some logarithmically-enhanced radiative corrections, such as the anapole moment effects, cannot be cannot be eliminated by the RGE for 
$\sin^2\hat\theta_W(\mu)$.}. As we discuss below, the evolution of $\sin^2\hat\theta_W(\mu)$ can be related in a straightforward way to that of ${\hat\alpha}(\mu)$, the QED coupling in the  $\overline{\rm MS}$-scheme that has been thoroughly studied elsewhere. Doing so allows us to draw upon known results for the QED $\beta$-function and -- in conjunction with suitable matching conditions --  to improve the precision of the Standard Model 
predictions for low-energy observables by incorporating various higher order effects. Indeed, although 
the one-loop RGE for $\sin^2\hat\theta_W(\mu)$ has been well-studied by others,
one emphasis of the present work is the inclusion of higher order QED and 
perturbative QCD contributions in a systematic way. We discuss the RGE in 
Sections~\ref{sec:rgeLO} and~\ref{sec:rgeHO}, and matching conditions in 
Section~\ref{sec:matching}. 

A major complication arises when the contribution of the light quark flavors is
considered for $\mu$ of order a hadronic scale, $\Lambda$, or smaller. 
In this regime, QCD corrections to the RGE ($\beta$-function) cannot be 
obtained using perturbative methods. An analogous problem is well-known 
to arise for $\hat\alpha(\mu)$ when its value is desired at scales 
similar to or greater than $\Lambda$. We address this problem in 
Sections~\ref{sec:hadronic} and~\ref{sec:uncertainties}, as well as 
two appendices. We argue that the corresponding theoretical uncertainty is well
below the anticipated experimental errors and provide a new estimate of 
this uncertainty that is substantially smaller in magnitude than 
the previously-quoted one~\cite{Marciano:1993ep}. Non-logarithmic contributions and some more formal aspects are discussed in Section~\ref{sec:finite} while numerical results and a plot of 
$\sin^2\hat\theta_W(\mu)$ for $\mu=\sqrt{|q^2|}$ -- along with conclusions -- are presented in Section~\ref{sec:conclude}.  We note there our results may also be applied to the recent
studies of deep-inelastic neutrino-nucleus scattering carried out by the NuTeV 
Collaboration~\cite{Zeller:2001hh}.

\section{Leading order RGE analysis}
\label{sec:rgeLO}
The quantity $\sin^2\hat\theta_W(0)$ is related to 
$\hat{s}^2_Z \equiv \sin^2\hat\theta_W(M_Z)$ by,
\be
   \sin^2\hat\theta_W(0) \equiv \hat\kappa(0) \sin^2\hat\theta_W(M_Z)
   \equiv [ 1 + \Delta\hat\kappa(0)] \hat{s}^2_Z,
\label{sin2def}
\ee 
where $\Delta\hat\kappa(0)$ is a universal (flavor-independent) radiative 
correction. In this Section we are interested in logarithmic contributions to 
$\hat\kappa(0)$ of the form, 
$$ 
   {\hat\alpha\over\pi\sin^2\hat\theta_W} \ln {M_Z^2\over m_i^2},
$$ 
and in particular in the scale that should be used in $\alphat$ and 
$\sin^2\hat\theta_W$, appropriate for re-summing the leading, large logarithms to all 
orders. These logarithms arise from scale dependent self-energy mixing diagrams
where one external leg is a photon and the other one is a $Z$-boson. Thus, 
$v_f$ acquires a compensating scale dependence,
\be
   \hat{v}_f \left( \sqrt{\mu^2 + \Delta\mu^2} \right) = \hat{v}_f(\mu) + 
   {\hat\alpha(\mu)\over 24\pi} Q_f \sum\limits_i [N^c_i \gamma_i v_i(\mu) Q_i]
   \ln {\mu^2 + \Delta\mu^2\over\mu^2},
\ee
where $N^c_q = N_C = 3$ (for quarks) and $N^c_\ell = 1$ (for leptons) is 
the color factor. We have written the sum in general form to also allow chiral 
fermions and bosonic degrees of freedom (to put the $W^\pm$ and Higgs ghosts on
the same footing as the fermions and to facilitate the discussion in 
Section~\ref{sec:finite}), where the spin dependent factors $\gamma_i$ are 
shown in Table~\ref{gamma}.
\begin{table}[t]
\centering
\begin{tabular}{|l|r|}
\hline
 field & $\gamma_i$ \\
\hline\hline
real scalar & 1 \\
complex scalar & 2 \\
chiral fermion & 4 \\
Majorana fermion & 4 \\
Dirac fermion & 8 \\
massless gauge boson & $-22$ \\
\hline
\end{tabular}
\caption{Weight factors $\gamma_i$ entering the leading RGE coefficients for
the weak mixing angle.}
\label{gamma}
\end{table}
With these conventions, the factor of 1/24 also appears in the lowest order QED
$\beta$-function coefficient,
\be 
   \mu^2 {d \hat\alpha\over d\mu^2} = 
   {\hat\alpha^2\over 24\pi} \sum\limits_i N^c_i \gamma_i Q_i^2.
\label{RGEQED}
\ee
This implies the RGE,
\be 
   \mu^2 {d \hat{v}_f\over d\mu^2} = {\hat\alpha\over 24\pi} Q_f
   \sum\limits_i N^c_i \gamma_i \hat{v}_i Q_i,
\ee
or in terms of the variable $\hat{X} = \sum_i N^c_i \gamma_i \hat{v}_i Q_i$,
\be
   {d \hat{X}\over \hat{X}} = {\hat\alpha\over 24\pi} {d \mu^2\over\mu^2}
   \sum\limits_i N^c_i \gamma_i Q_i^2 = {d \hat\alpha\over\hat\alpha},
\label{RGELO}
\ee
where in the second equality we used Eq.~(\ref{RGEQED}). This is solved by,
\be
   \sin^2\hat\theta_W(\mu) = {\hat\alpha(\mu)\over\hat\alpha(\mu_0)} 
   \sin^2\hat\theta_W(\mu_0) + 
   {\sum_i N^c_i \gamma_i Q_i T_i\over \sum_i N^c_i \gamma_i Q_i^2} 
   \left[ 1 - {\hat\alpha(\mu)\over\hat\alpha(\mu_0)} \right],
\label{solutionLO}
\ee
or using the explicit solution to the one-loop RGE in Eq.~(\ref{RGEQED}) we
obtain the simpler form,
\be
   \sin^2\hat\theta_W(\mu) = \sin^2\hat\theta_W(\mu_0)\left[ 1 + 
   {\hat\alpha(\mu) \over 24\pi\sin^2\hat\theta_W(\mu_0)} \sum\limits_i N_i^c
   \gamma_i Q_i [T_i - Q_i\sin^2\hat\theta_W(\mu_0)] \ln{\mu_0^2\over \mu^2} 
   \right].
\label{solutionLO2}
\ee
The result in Eq. (\ref{solutionLO2}) re-sums all logarithms of ${\cal O}(\alpha^n\ln^n{\mu_0\over\mu})$ 
provided there is no particle threshold between $\mu$ and $\mu_0$. To avoid 
reintroduction of spurious logarithms this solution must be applied 
successively from one particle threshold to the next. Crossing a threshold from
above, the corresponding particle is integrated out, and one continues with 
an effective field theory without this particle. In contrast, changing from one
effective field theory to another far away from the physical mass of 
the particle would not formally affect the truncated one-loop result, but it 
would spoil its re-summation. 

Eq.~(\ref{solutionLO2}) applied to $\Delta\kappa(0)$ can be brought into 
the well-known form~\cite{Marciano:1980pb},
\be
   \Delta\kappa(0) = {\alpha\over\pi\hat{s}^2_Z} \left[ {1\over 6}\sum\limits_f
   N_f^c Q_f (T_f - 2 Q_f \hat{s}^2_Z) \ln{M_Z^2\over m_f^2} - \left( 
   {43\over 24} - {7\over 4} \hat{s}^2_Z \right) \ln {M_Z^2\over M_W^2}\right],
\label{kappaLO}
\ee
where the sum is over all SM Dirac fermions excluding the top quark. The last 
term is the $W^\pm$ contribution with its coefficient obtainable from 
Table~\ref{gamma} when a pair of massless gauge bosons ($T_i = \pm 1$) is 
combined with a complex scalar Goldstone degree of freedom ($T_i = \pm 1/2$).

\section{Higher order RGE analysis}
\label{sec:rgeHO}
In this Section we will generalize the leading order analysis and re-sum 
next-to-leading (NL) logarithms of ${\cal O}(\alpha^{n+1}\ln^n{\mu_0\over\mu})$
and ${\cal O}(\alpha\alpha_s^n\ln^n{\mu_0\over\mu})$, as well as the NNL 
logarithms of ${\cal O}(\alpha\alpha_s^{n+1}\ln^n{\mu_0\over\mu})$, and 
the NNNL logarithms of ${\cal O}(\alpha\alpha_s^{n+2}\ln^n{\mu_0\over\mu})$.
The leading order RGE~(\ref{RGELO}) supplemented by terms of 
${\cal O}(\alpha^2)$, ${\cal O}(\alpha\alpha_s)$, ${\cal O}(\alpha\alpha_s^2)$,
and ${\cal O}(\alpha\alpha_s^3)$ reads,
\be 
   \mu^2 {d \hat{v}_f\over d\mu^2} = {\hat\alpha\over 24\pi} Q_f \left[ 
   \sum\limits_i K_i \gamma_i \hat{v}_i Q_i + 12 \sigma \left( 
   \sum\limits_q \hat{v}_q \right) \left( \sum\limits_q Q_q \right) \right],
\label{RGEHO}
\ee
where in the case of quarks ($n_q$ is the effective number of quarks),
\be
   K_i = N_i^c \left\{ 1 + {3\over 4} Q_i^2 {\hat\alpha\over\pi} + 
  {\hat\alpha_s\over\pi} + {\hat\alpha_s^2\over\pi^2} \left( {125\over 48} - 
  {11\over 72} n_q \right) + \right.
\label{ki}
\ee
$$ \left. {\hat\alpha_s^3\over\pi^3} \left[ {10487\over 1728} + 
   {55\over 18}\zeta(3) - \left( {707\over 864} + {55\over 54}\zeta(3) \right)
   n_q - {77\over 3888} n_q^2 \right] \right\},
$$
contains QED and QCD corrections~\citer{Gorishnii:1988bc,Larin:1994va} to 
the lowest order (non-singlet) vacuum polarization diagrams. For leptons only 
the term involving $\hat\alpha$ is kept, while for bosons we restrict ourselves
to the lowest order $\beta$-function\footnote{We do so because full two-loop 
electroweak calculations are generally incomplete and therefore only 
the leading order electroweak terms included in most current definitions of 
$\overline{\rm MS}$ quantities. Moreover, the structure of the RGE would change
relative to Eq.~(\ref{RGEHO4}) below, spoiling the corresponding 
solution~(\ref{solutionHO}).  Because the logarithms, $\ln M_Z/M_W$, are not 
large, neglecting them in electroweak two-loop terms is numerically 
insignificant.}, {\em i.e.\/} $K_{W^\pm} = 1$. The second sum in 
Eq.~(\ref{RGEHO}) is over Dirac quark fields, and
\be
   \sigma = {\hat\alpha_s^3\over\pi^3} \left[ {55 \over 216} - 
   {5 \over 9} \zeta(3) \right] + {\cal O}(\hat\alpha_s^4),
\label{sigma}
\ee
parametrizes the QCD singlet contribution. In a singlet (QCD annihilation) 
diagram two independent fermion loops are attached to the $\gamma$ and $Z$ and 
connected to each other by gluons or photons. Due to Furry's theorem, 
connections containing a photon first arise at ${\cal O}(\alpha^2\alpha_s^2)$ 
and can safely be neglected. Defining $\hat{s}^2 = \sin^2\hat\theta_W(\mu)$ we 
rewrite Eq.~(\ref{RGEHO}),
\be 
   \mu^2 {d \hat{s}^2 \over d\mu^2} = {\hat\alpha\over\pi} 
   \left[ {1\over 24} \sum\limits_i K_i \gamma_i (Q_i^2 \hat{s}^2 - T_i Q_i) +
   \sigma \hat{s}^2 \left( \sum\limits_q Q_q \right)^2 - {\sigma\over 2}
   \left( \sum\limits_q T_q \right) \left( \sum\limits_q Q_q \right) \right].
\label{RGEHO2}
\ee
Similarly, the RGE for $\hat\alpha$ including higher orders reads,
\be 
   \mu^2 {d \hat\alpha\over d\mu^2} = {\hat\alpha^2\over\pi} 
   \left[ {1\over 24} \sum\limits_i K_i \gamma_i Q_i^2 + 
   \sigma \left( \sum\limits_q Q_q \right)^2 \right],
\label{RGEQEDHO}
\ee
and we obtain,
\be 
   \mu^2 {d \over d\mu^2} \left( {\hat{s}^2\over\hat\alpha} \right) = 
   - {1\over 24\pi} \sum_i K_i \gamma_i T_i Q_i - {\sigma\over 2\pi} 
   \left( \sum\limits_q T_q \right) \left( \sum\limits_q Q_q \right).
\label{RGEHO3}
\ee
To facilitate the integration and to relate hadronic contributions as far as
possible to the ones in $\hat\alpha$, we use Eq.~(\ref{RGEQEDHO}) again and
eliminate all $\hat{\alpha}_s$ dependent terms in Eq.~(\ref{ki}). With
the coefficients\footnote{The explicit factor of 1/2 in $\lambda_1$ compared
to the coefficient in the leading order solution~(\ref{solutionLO}) arises 
because the electric charges in the denominator of the latter are summed over
left and right chiralities while only left chiralities appear in 
the numerator.},
\be
   \lambda_1 = {\sum_q T_q Q_q\over 2 \sum_q Q^2_q}, \hspace{50pt} \lambda_2 = 
   {1\over 8} \sum\limits_{i\neq q} \gamma_i (\lambda_1 Q_i^2 - T_i Q_i) =
   {1\over 8} \sum\limits_i N_i^c \gamma_i (\lambda_1 Q_i^2 - T_i Q_i),
\label{lambda12}
\ee
and,
\be
   \lambda_3 = {\sum_i N_i^c \gamma_i [ \lambda_1 Q_i^4 - T_i Q_i^3 ] \over
   \sum_i N^c_i \gamma_i Q^2_i}, \hspace{40pt}
   \lambda_4 = \left[ \lambda_1 \left( \sum\limits_q Q_q\right)^2 - {1\over 2}
   \left( \sum\limits_q T_q \right) \left( \sum\limits_q Q_q \right) \right],
\label{lambda34}
\ee
shown in Table~\ref{coeff} this can be brought into the form,
\begin{table}[t]
\centering
\begin{tabular}{|r|c|c|c|c|}
\hline
 energy range & $\lambda_1$ & $\lambda_2$ & $\lambda_3$ & $\lambda_4$  \\
\hline\hline
$\bar{m}_t\leq \mu            \hspace*{36pt}$&  9/20 & 289/80  & 14/55 & 9/20\\
$M_W      \leq \mu < \bar{m}_t\hspace*{6pt}$ & 21/44 & 625/176 &  6/11 & 3/22\\
$\bar{m}_b\leq \mu < M_W      \hspace*{0pt}$ & 21/44 &  15/22  & 51/440& 3/22\\
$m_\tau   \leq \mu < \bar{m}_b\hspace*{6pt}$ &  9/20 &   3/5   &  2/19 & 1/5 \\
$\bar{m}_c\leq \mu < m_\tau   \hspace*{5pt}$ &  9/20 &   2/5   &  7/80 & 1/5 \\
$\bar{m}_s\leq \mu < \bar{m}_c\hspace*{5pt}$ &  1/2  &   1/2   &  5/36 &  0  \\
$\bar{m}_d\leq \mu < \bar{m}_s\hspace*{5pt}$ &  9/20 &   2/5   & 13/110& 1/20\\
$\bar{m}_u\leq \mu < \bar{m}_d\hspace*{5pt}$ &  3/8  &   1/4   &  3/40 &  0  \\
$m_\mu    \leq \mu < \bar{m}_u\hspace*{5pt}$ &  1/4  &    0    &   0   &  0  \\
$m_e      \leq \mu < m_\mu    \hspace*{5pt}$ &  1/4  &    0    &   0   &  0  \\
$              \mu < m_e      \hspace*{6pt}$ &   0   &    0    &   0   &  0  \\
\hline
\end{tabular}
\caption{Coefficients entering the higher order RGE for the weak mixing angle.
For the definition of quark threshold masses, $\bar{m}_q$, see 
Section~\ref{sec:hadronic}. Below hadronic scales $\lambda_1$ is not defined
through Eq.~(\ref{lambda12}) and can be chosen arbitrarily; we chose the value,
$\lambda_1 = 1/4$, to obtain $\lambda_2 = \lambda_3 = 0$.}
\label{coeff}
\end{table}
\be
   \mu^2 {d \over d\mu^2} \left( {\hat{s}^2 - \lambda_1\over\hat\alpha} - 
   {3 \lambda_3\over 4\pi} \ln\hat\alpha + {\tilde\sigma\over\pi} \right) = 
   {\lambda_2\over 3\pi}.
\label{RGEHO4}
\ee
For the last term on the left hand side we have used the lowest order QCD 
$\beta$-function coefficient and have defined
\be
   \tilde\sigma(\mu) = \lambda_4 {\hat\alpha_s^2(\mu)\over\pi^2} {5\over 36} 
   {11 - 24\zeta(3) \over 33 - 2 n_q} + {\cal O}(\hat\alpha_s^3).
\label{sigmatilde}
\ee
The solution to Eq.~(\ref{RGEHO4}) is given by,
\be
   \hat{s}^2(\mu) = {\hat\alpha(\mu)\over\hat\alpha(\mu_0)} \hat{s}^2(\mu_0) + 
   \lambda_1 \left[ 1 - {\hat\alpha(\mu)\over\hat\alpha(\mu_0)} \right] + 
   {\hat\alpha(\mu)\over\pi} \left[ {\lambda_2\over 3}\ln {\mu^2\over\mu_0^2} +
   {3 \lambda_3 \over 4}\ln {\hat\alpha(\mu)\over\hat\alpha(\mu_0)} +
   \tilde\sigma(\mu_0) - \tilde\sigma(\mu) \right].
\label{solutionHO}
\ee
Hadronic uncertainties are induced through $\hat\alpha(\mu)$, through 
the {\em relative\/} values of the light quark threshold masses, $\bar{m}_u$, 
$\bar{m}_d$, and $\bar{m}_s$ (these are needed as they determine the change of
the coefficients $\lambda_i$ according to Table~\ref{coeff}), and through the 
singlet contribution proportional to $\tilde\sigma(\mu_0) - \tilde\sigma(\mu)$.

\section{Matching conditions}
\label{sec:matching}
At the threshold of fermion $f$ we find,
\be
   \sin^2\hat\theta_W(m_i)^- = {\hat\alpha(m_i)^-\over \hat\alpha(m_i)^+} 
   \sin^2\hat\theta_W(m_i)^+ + {Q_i T_i \over 2 Q_i^2}
   \left[ 1 - {\hat\alpha(m_i)^- \over \hat\alpha(m_i)^+} \right],
\label{threshold}
\ee
where the plus (minus) superscript denote the effective theory including
(excluding) fermion $f$. While the one and two-loop $\beta$-function 
coefficients are well-known to be renormalization scheme independent, 
the matching conditions are renormalization scheme and even regularization 
scheme dependent.  The $\overline{\rm MS}$-scheme is defined by dimensional 
regularization which generates no ${\cal O}(\alpha)$ matching terms for 
scalars; with the usual additional requirement\footnote{Other definitions do 
occur in the literature, however.} that the Clifford algebra is kept in four 
dimensions the same holds for spin-1/2 fermions. We include the RGE matching 
conditions for $\hat\alpha$ at the threshold for fermion $f$ at the orders 
$\alpha^2$, $\alpha\alpha_s$, and 
$\alpha\alpha_s^2$~\cite{Chetyrkin:1996cf,Erler:1998sy},
\be
   {1\over\hat\alpha^+(m_f)} = {1\over\hat\alpha^-(m_f)} - {Q_f^2\over\pi} 
   \left\{ {15\over 16} N^c_f Q_f^2 {\hat\alpha(m_f)\over\pi} + \right.
\label{matching}
\ee
$$ \left. {(N^c_f - 1)\over 2}{\hat\alpha_s(m_f)\over\pi} \left[ {13\over 12} +
   {\hat\alpha_s(m_f)\over\pi} \left( {655\over 144} \zeta(3) - {3847\over 864}
   + {361\over 1296} n_q + {295\over 1296} {\sum_{q \neq f} Q_q^2\over Q_f^2} 
   \right) \right] \right\} . $$
Here, $n_q$ is the number of quarks including the threshold quark\footnote{In 
writing this equation we assume that $m_f$ is an $\overline{\rm MS}$-mass 
(which is free of renormalon ambiguities and assures a better convergence of 
the perturbative series) to the extent to which QCD effects are concerned, but 
a pole mass for both leptons and quarks with respect to QED (to comply with 
standard conventions in the literature). This results in a somewhat awkward 
definition for quarks but is of no importance in practice since 
the ${\cal O}(\alpha^2)$ corrections are very small.} $f$. 
Eq.~(\ref{threshold}) will then induce the corresponding matching contributions
to the weak mixing angle. 

In contrast to fermions and scalars, gauge bosons induce an ${\cal O}(\alpha)$
threshold shift~\cite{Hall:1980kf},
\be
   {1\over \alpha_i^+} = {1\over \alpha_i^-} + {C(R)\over 12\pi},
\ee
where $C(R)$ is the quadratic Casimir of the (in general reducible) gauge boson
representation, normalized such that, {\em e.g.}, 
$C({\rm adjoint\/}[SU(N)]) = N$. Thus, integrating out the $W^\pm$ bosons 
induces a shift\footnote{This shift is an artifact of using modified minimal 
subtraction in dimensional regularization. It is precisely canceled against 
a conversion constant~\cite{Antoniadis:1982vr,Langacker:1992rq} which relates 
the $\overline{\rm MS}$-scheme to the $\overline{\rm DR}$-scheme. The latter is
defined by dimensional reduction and is used in supersymmetric theories.} in 
the electromagnetic coupling,
\be
   {1\over \hat\alpha (M_W)^+} = {1\over \hat\alpha (M_W)^-} + {1\over 6\pi},
\label{eq:thresholda}
\ee
and (generalizing Eq.~(\ref{threshold}) appropriately) in the weak mixing 
angle,
\be
   \sin^2\hat\theta_W (M_W)^+ = 
   1 - {\hat\alpha (M_W)^+\over \hat\alpha (M_W)^-} \cos^2\hat\theta_W (M_W)^-.
\label{eq:thresholdb}
\ee

\section{Hadronic contribution}
\label{sec:hadronic}
The ambiguity in the values of the three light quark masses plus the singlet 
contribution to Eq.~(\ref{solutionHO}) introduce four sources of hadronic 
uncertainties in $\sin^2\hat\theta_W(0)$. This problem is familiar from 
the evaluation of $\alpha(M_Z)$, where it can be addressed by relating it 
{\em via\/} a dispersion relation to $e^+ e^- \to$ hadrons cross section data, 
or (using in addition isospin symmetry) to $\tau$ decay spectral functions. 
The same strategy could in principle be applied here, except that 
the experimental information would have to be separated in charge 2/3 ($u$) 
{\em vs.\/} charge $-1/3$ ($d$ and $s$) quarks, or assuming isospin symmetry, 
in $s$ {\em vs.\/} the first generation quarks. This is a difficult task: 
{\it e.g.,} a $K^\pm$ final state in $e^+ e^-$ annihilation can be produced 
directly through an $\bar{s}\gamma_\mu s$ current or by splitting of a gluon 
radiated off a quark originating from 
a $\bar{u}\gamma_\mu u + \bar{d}\gamma_\mu d$ (isoscalar) current. It would be
even more difficult to isolate the singlet contribution. As for $\tau^-$ 
decays, the relatively large $m_s$ induces a sizable axial-vector contribution 
to final states with strangeness, $S= -1$. At least presently, it cannot be 
cleanly separated from the vector contribution which is the relevant one for 
the problem at hand. We will therefore follow a different strategy and show 
that the four uncertainties can be traded for the uncertainties associated with
(i) the value of ${\hat\alpha}(M_Z)$, which we denote $\delta_\alpha\sin^2\hat\theta_W(0)$, with (ii) 
the separation of strange and first generation quark effects, indicated by
$\delta_s\sin^2\hat\theta_W(0)$, with (iii) deviations from isospin symmetry, 
$\delta_{\rm CVC}\sin^2\hat\theta_W(0)$, and with (iv) Zweig (OZI) rule 
deviations, $\delta_{\rm OZI}\sin^2\hat\theta_W(0)$.

As discussed in Section~\ref{sec:rgeHO}, the contributions of leptons and heavy
quarks can be computed unambiguously in perturbation theory. Indeed, using QCD
$\overline{\rm MS}$ 
masses\addtocounter{footnote}{-2}\footnotemark\addtocounter{footnote}{1}, 
$\hat{m}_q (\hat{m}_q)$, will provide a small truncation error, 
$\delta_{\rm PQCD}\sin^2\hat\theta_W(0)$, in the perturbative expansion (cf.\ 
the small matching coefficients in Eq.~(\ref{matching})). Moreover, 
the numerical values of $\hat{m}_q(\hat{m}_q)$ can be determined to sufficient
precision that is {\em not\/} limited by uncertainties of the order of 
hadronic scales. Therefore, we compute first $\sin^2\hat\theta_W(\bar{\mu})$
from $\sin^2\hat\theta_W(M_Z)$ (which can be taken from $Z$ pole experiments), 
where $\bar{\mu}$ corresponds to a scale where the heavy flavors ($b$, $c$, and
$\tau$) are integrated out, and at which we still have sufficient confidence in
the convergence of perturbative QCD ({\em i.e.\/}, of order 1~GeV). 

We constrain the contributions of light quarks to 
$\sin^2\hat\theta_W(\bar{\mu})$ phenomenologically. Our strategy is to employ 
the $u$, $d$, and $s$ quark contributions to $\hat\alpha(\bar{\mu})$, 
$\Delta\hat\alpha^{(3)}(\bar{\mu})$, as a constraint, and to find upper and 
lower bounds on the strange quark contribution relative to the contributions of
the first generation quarks. In the following we assume isospin symmetry and 
a vanishing singlet contribution. Deviations from these assumptions will be 
addressed in Section~\ref{sec:uncertainties}.

To facilitate the discussion, we will adopt {\em definitions\/} of threshold 
quark masses, $\bar{m}_u$, $\bar{m}_d$, and $\bar{m}_s$, such that 
Eq.~(\ref{solutionHO}) remains valid with trivial matching conditions, 
$\alpha_i^+(\bar{m}_q) = \alpha_i^-(\bar{m}_q)$. Thus, $\bar{m}_u$, 
$\bar{m}_d$, and $\bar{m}_s$ define the ranges in Table~\ref{coeff}. Their 
values can be constrained phenomenologically\footnote{See 
Ref.~\cite{Jegerlehner:1991dq} for an earlier determination.}, but their 
relation to other mass definitions, such as constituent masses or current 
masses, cannot be written down in a perturbative sense. One combination of 
the three light quark threshold masses is constrained to reproduce 
$\hat\alpha(\bar{\mu})/\alpha$.  If we assume isospin symmetry, 
$\bar{m}_u = \bar{m}_d$, and a vanishing singlet contribution, we have only one
unknown parameter, say $\bar{m}_s$, to describe 
$\delta_s\sin^2\hat\theta_W(0)$. Before we use physics arguments to constrain 
$\bar{m}_s$, we compute $\bar{m}_c$ and $\bar{m}_b$ perturbatively to gauge 
the behavior of heavy quarks.  To order $\alpha_s^2$ we 
have~\cite{Chetyrkin:1996cf,Erler:1998sy},
\be
   \ln {\mu^2\over \hat{m}^2(\mu)} + {\hat\alpha_s(\mu)\over\pi} 
   \left[ {13\over 12} - \ln {\mu^2\over \hat{m}^2(\mu)} \right] + 
   {\hat\alpha_s^2\over \pi^2} \left[ {655\over 144}\zeta(3) - {3847\over 864} 
   + n_q {361\over 1296} + {295\over 1296} {\sum_{q \neq f} Q_q^2\over Q_f^2} 
   \right] = 0,
\ee
which implies,
\be
   \bar{m} = \hat{m}(\hat{m}) \exp\left[ - {13\over 24} {\hat\alpha_s(\hat{m})
   \over\hat\alpha_s(\hat{m}) + \pi} - {\hat\alpha_s^2\over 288\pi^2} \left( 
   655\zeta(3) - {3847\over 6} + {361\over 9} n_q + 
   {295\over 9} {\sum_{q \neq f} Q_q^2\over Q_f^2} \right) \right].
\label{barmc}
\ee
With the input values (obtained from a global fit to precision data),
$\hat\alpha_s (M_Z) = 0.1214 \pm 0.0018$, 
$\hat{m}_c (\hat{m}_c) = 1.285_{-0.047}^{+ 0.040}$, and
$\hat{m}_b (\hat{m}_b) = 4.205 \pm 0.031$, we find $\bar{m}_c = 1.176$~GeV and
$\bar{m}_b = 3.995$~GeV.

\subsubsection*{The heavy $\bar{m}_s$ limit}
To obtain a lower limit on the strange quark contribution to 
$\hat\alpha(\bar\mu)$, we consider the case in which the strange quark is 
assumed to behave like a heavy quark. In this case, $\bar{m}_s$ would be 
related to $M_\phi$ in a similar way as $\bar{m}_c$ (or $\hat{m}_c(\hat{m}_c)$)
is related\footnote{More precisely, QCD sum rules relate 
$\bar{m}_c$ ($\bar{m}_b$) rigorously to a weighted sum over $\Psi$ ($\Upsilon$)
resonances plus a continuum contribution. For the present consideration we 
restrict ourselves to the lowest lying resonance which carries the largest 
weight.} to $M_{J/\Psi}$. Defining $\xi_q = 2 \bar{m}_q/M_{1S}$ where $M_{1S}$ 
is the mass of the $1S$ $q\bar{q}$ resonance, we have that asymptotically 
$\xi_q \to 1$ for $\bar{m}_q \to \infty$ and $\xi_q \to 0$ for $\bar{m}_q\to 0$
(in the chiral limit the quark contribution is logarithmically divergent). 
Thus, for a heavy quark $\xi_q \sim 1$, while for a light quark $\xi_q \ll 1$.
Also, we expect $\bar{m}_1 < \bar{m}_2 \Longrightarrow \xi_1 < \xi_2$.  
As an illustration, with the numerical values of $\bar{m}_c$ and $\bar{m}_b$ 
from above we obtain $\xi_c = 0.759$, $\xi_b = 0.845$, and 
$\bar{m}_s = \xi_s M_\phi/2 < \xi_c M_\phi/2 = 387$~MeV. As a refinement we
introduce scale dependent QCD correction factors, 
$K_{\rm QCD}^q = K_{\rm QCD}(\bar{\mu},\bar{m}_q)$, where 
$K_{\rm QCD}(\bar{\mu}_1,\bar{\mu}_2)$ denotes the average QCD correction to 
the QED $\beta$-functions for RGE running between scales 
$\bar{\mu}_1$ and $\bar{\mu}_2$. One thus expects
$\bar{m}_1 < \bar{m}_2 \Longrightarrow K_{\rm QCD}^1 > K_{\rm QCD}^2$. Since 
Eq.~(\ref{barmc}) applied to $m_c$ still shows satisfactory convergence, we can
safely choose $\bar{\mu} = \bar{m}_c$, 
\be
  \Delta_s\hat\alpha(\bar{m}_c) =
  Q_s^2 {\alpha\over\pi} K_{\rm QCD}^s \ln {\bar{m}_c^2\over\bar{m}_s^2} >
  Q_s^2 {\alpha\over\pi} K_{\rm QCD}^c \ln {\bar{m}_c^2\over\bar{m}_s^2} >
  {2\alpha\over 9\pi}    K_{\rm QCD}^c \ln {M_{J/\Psi}\over M_\phi} =
  6.9 \times 10^{-4}.
\label{lowerlimit}
\ee
For the numerical evaluation we have used the QCD correction in Eq.~(\ref{ki})
applied to the effective theory with $n_q = 3$ quarks,
\be
  K_{\rm QCD}^s > K_{\rm QCD}^c = 1 + {\hat\alpha_s(\bar{m}_c)\over\pi} +
  {103\over 48} {\hat\alpha_s^2(\bar{m}_c)\over\pi^2} + 
  {1979\over 576} {\hat\alpha_s^3(\bar{m}_c)\over\pi^3} = 1.209,
\ee
and we have used $\hat\alpha_s(\bar{m}_c) = 0.469$, again corresponding to 
$\hat\alpha_s(M_Z) = 0.1214$.

\subsubsection*{The $SU(3)$ limit}
Since $m_s > m_d \gsim m_u$ at any scale and in any reasonable 
definition\footnote{This statement holds because small non-universal mass 
renormalization corrections from QED and the electroweak interactions can be 
neglected.} and scheme, we conclude that the $SU(3)$ symmetric case,
$\bar{m}_u = \bar{m}_d = \bar{m}_s$, implies an upper limit on the relative 
strange quark contribution to $\hat\alpha(\bar{\mu})$,
\be
  \Delta_s\hat\alpha(\bar{\mu}) \leq
  {1\over 6} \left[ \Delta_u\hat\alpha(\bar{\mu}) + 
  \Delta_d\hat\alpha(\bar{\mu}) + \Delta_s\hat\alpha(\bar{\mu}) \right] \equiv
  {1\over 6} \Delta\hat\alpha^{(3)}(\bar{\mu}).
\label{equalm}
\ee
This crude limit can be strengthened by considering the phenomenological 
constraint,
\be
   {2 \alpha\over\pi} \left[ 
   (Q^2_u + Q^2_d) K_{\rm QCD}^{u,d} \ln {2\bar{\mu}\over\xi_{u,d} M_\omega}
          + Q^2_s  K_{\rm QCD}^s     \ln {2\bar{\mu}\over\xi_s     M_\phi} 
   \right] = \Delta\hat\alpha^{(3)} (\bar{\mu}),
\label{pheno}
\ee
and by imposing $SU(3)$ symmetry through $\xi_u = \xi_d = \xi_s$ and 
$K_{\rm QCD}^{u} = K_{\rm QCD}^d = K_{\rm QCD}^s$. This maximizes the ratio of
the strange quark contribution to the one of the first generation quarks, 
\be
   {\Delta_s\hat\alpha(\bar{\mu})\over\Delta_{u+d}\hat\alpha(\bar{\mu})}
   = {Q^2_s\over Q^2_u + Q^2_d} \left[ 1 + {K^s_{\rm QCD} \ln {\xi_{u,d} \over
   \xi_s} - K^s_{\rm QCD} \ln {M_\phi \over M_\omega} - (K^{u,d}_{\rm QCD} - 
   K^s_{\rm QCD}) \ln {2\bar{\mu} \over \xi_{u,d} M_\omega}\over
   K^{u,d}_{\rm QCD} \ln {2\bar{\mu} \over \xi_{u,d} M_\omega}} \right].
\ee
All three corrections terms to the charge square ratio are indeed negative, and
we used the second one as an improvement. In the limit $\xi_s \to 0$, 
Eq.~(\ref{pheno}) would reproduce relation~(\ref{equalm}) but now we have 
the constraint,
\be
   \xi_s >
   {2\bar{\mu}\over M_\omega^{5/6} M_\phi^{1/6}} \exp \left[ - {3\pi\over 4} 
   {\Delta\hat\alpha^{(3)}(\bar{\mu})\over\alpha K_{\rm QCD}^s} \right].
\label{xis}
\ee
For the numerical evaluation we convert the contribution to the on-shell 
definition of $\alpha(M_Z)$, $\Delta\alpha_{\rm had}(M_Z) = 0.00577\pm 
0.00010$~\cite{Davier:1998si,Davier:2003pw}, from the energy range up to 
1.8~GeV, to the $\overline{\rm MS}$-scheme,
$$\Delta\hat\alpha^{(3)}(\mbox{1.8~GeV}) = \Delta\alpha_{\rm had}(M_Z) +
  {2\alpha\over3\pi}\left[ {5\over 3} + \left( {\hat\alpha_s\over\pi} +
  {\hat\alpha\over 4\pi} \right) \left( {55\over 12} - 4 \zeta(3) + 
  {2\hat{m}_s^2\over(1.8~{\rm GeV})^2} \right) + \right. $$
$$\left. {\hat\alpha_s^2\over\pi^2}\left( {34525\over 864} - {9\over 4}\zeta(2)
  - {715\over 18}\zeta(3) + {25\over 3}\zeta(5) + F(\hat{m}_c,\hat{m}_b) 
  \right) \right] = 0.00831 \pm 0.00010,$$
where all $\overline{\rm MS}$ running couplings and masses are to be taken at 
$\mu = 1.8$~GeV. $F(\hat{m}_c,\hat{m}_b)$ contains decoupling charm and bottom
mass effects~\cite{Larin:1994va,Kniehl:1989kz,Hoang:1994it}. We choose again 
$\bar{\mu} = \bar{m}_c$ and use the 4-loop RGE to obtain,
\be
   \Delta\hat\alpha^{(3)}(\bar{m}_c) = 0.00678 \pm 0.00010,
\label{deltalp}
\ee
which using the $SU(3)$ bound~(\ref{xis}) corresponds to $\xi_s > 0.470$ and 
$\bar{m}_s > 240$~MeV. Inserting Eq.~(\ref{xis}) into Eq.~(\ref{pheno}) in 
the $SU(3)$ limit yields\footnote{Exact $SU(3)$ symmetry would imply 
$M_\rho = M_\omega$; since we are interested in an upper limit on the strange 
quark contribution we have choose the (larger) phenomenological value of 
$M_\omega$.},
\be
  \Delta_s\hat\alpha(\bar{m}_c) <
  {\Delta\hat\alpha^{(3)}(\bar{m}_c)\over 6} - {5\over 27} {\alpha\over\pi} 
  K_{\rm QCD}^s \ln {M_\phi\over M_\omega} < 
  {\Delta\hat\alpha^{(3)}(\bar{m}_c)\over 6} - {5\over 27} {\alpha\over\pi} 
  K_{\rm QCD}^c \ln {M_\phi\over M_\omega} = 9.9 \times 10^{-4}.
\label{upperlimit}
\ee
We have assumed ideal $\omega-\phi$ mixing, {\em i.e.\/} that the $\phi$ 
resonance is a pure $s\bar{s}$ state. Allowing a non-ideal mixing angle, 
$\epsilon = 0.0548 \pm 0.0024 \neq 0$ (see Appendix~\ref{sec:epsilon}), shifts
the masses $M_\omega$ and $M_\phi$ to be used in Eqs.~(\ref{lowerlimit}) and 
(\ref{upperlimit}) by less than 1~MeV and yields a negligible effect. 

\subsubsection*{Implications}
From Eqs.~(\ref{lowerlimit}) and (\ref{upperlimit}) we conclude for 
the strange quark,
\be
   \Delta_s\hat\alpha(\bar{m}_c) = (8.4 \pm 1.5) \times 10^{-4}, \hspace{197pt}
   \bar{m}_s = 305_{+82}^{-65}\mbox{ MeV}.
\ee
and for the light quarks,
\be
   \Delta\hat\alpha^{(2)}(\bar{m}_s) = \Delta\hat\alpha^{(3)}(\bar{m}_c) -
   6 \Delta_s\hat\alpha(\bar{m}_c) = 0.00172 \mp 0.00090, \hspace{20pt}
   \bar{m}_d = \bar{m}_u = 176 \pm 9 \mbox{ MeV}.
\label{ud}
\ee
These results can be used in the master equation~(\ref{solutionHO}).  
As an illustration, the $SU(3)$ symmetric piece is well approximated by,
\be
   \hat{s}^2(\bar{m}_s) - \hat{s}^2(\bar{m}_c) = \left[ {1\over 2} - 
   \hat{s}^2(\bar{m}_c) \right] 6 \Delta_s\hat\alpha(\bar{m}_c) + 
   {2 \hat\alpha({\bar{m}_s})\over 3\pi} \left[ {1\over 4} - 
   \hat{s}^2(\bar{m}_c) \right] \ln {\bar{m}_c^2\over \bar{m}_s^2} 
   \left( 1 + {3 \hat\alpha({\bar{m}_s})\over 4\pi} \right),
\label{solutionuds}
\ee
which is obtained with the help of Table~\ref{coeff}, and where the last term 
is the leptonic ($e$ and $\mu$) contribution. Similarly, the $SU(3)$ breaking 
piece reads,
\be
   \hat{s}^2(\bar{m}_d) - \hat{s}^2(\bar{m}_s) = \left[ {9\over 20} - 
   \hat{s}^2(\bar{m}_s) \right] \Delta\hat\alpha^{(2)}(\bar{m}_s) + 
   {2 \hat\alpha({\bar{m}_d})\over 3\pi} \left[ {1\over 4} - 
   \hat{s}^2(\bar{m}_s) \right] \ln {\bar{m}_s^2\over \bar{m}_d^2}
   \left( 1 + {3 \hat\alpha({\bar{m}_d})\over 4\pi} \right),
\label{solutionud}
\ee
where we neglected the singlet piece involving $\bar\sigma$.

\section{Uncertainties}
\label{sec:uncertainties}
From Eq.~(\ref{deltalp}), as well as the sum of Eqs.~(\ref{solutionuds}) and
(\ref{solutionud}), we can bound the uncertainty induced by 
$\Delta\alpha^{(3)}(\bar{m}_c)$,
\be
   \delta_\alpha \sin^2\hat\theta_W(0) < \delta\Delta\alpha^{(3)}(\bar{m}_c)
   \left[ {1\over 2} - \hat{s}^2(\bar{m}_c) \right] = \pm 3 \times 10^{-5}.
\label{error1}
\ee
Similarly, from the first Eq.~(\ref{ud}) and from the comparison of 
the coefficients in Eqs.~(\ref{solutionuds}) and (\ref{solutionud}), we can 
estimate the uncertainty induced by $\Delta\alpha^{(2)}(\bar{m}_s)$,
\be
   \delta_s \sin^2\hat\theta_W(0) \approx 
   {1\over 20} \delta\Delta\alpha^{(2)}(\bar{m}_s) = \pm 5 \times 10^{-5}.
\label{error2}
\ee

\subsubsection*{The singlet contribution}
We obtained the theoretical bounds on $\bar{m}_s$ that are the basis for 
the error estimate~(\ref{error2}) by assuming isopsin symmetry and a vanishing 
singlet contribution. We now relax the latter assumption and allow OZI 
rule~\citer{Okubo:1963fa,Iizuka:1966fk} violation which leads to processes such
as $\phi\to\pi^0\gamma$ decays and which translates on a diagrammatic level to 
QCD annihilation (singlet) topologies. For charm and third generation quarks 
singlet contributions are tiny and easily included using Eq.~(\ref{sigma}) or 
Eq.~(\ref{sigmatilde}). We can then proceed with the effective theory 
containing only the three light quarks. Notice, that due to $Q_u + Q_d +Q_s=0$,
there is no singlet contribution (to the $\beta$-functions of neither $\alpha$ 
nor $\sin^2\theta_W$) in the limit of exact $SU(3)$ symmetry (see also 
the sixth entry for $\lambda_4$ in Table~\ref{coeff}). Moreover, allowing 
$SU(3)$ breaking effects --- but still working in the isospin symmetric limit 
--- will not directly affect the RGE~(\ref{RGEHO3}) because $T_u + T_d = 0$. 
The explicit singlet piece in Eq.~(\ref{solutionHO}) is an artifact of 
employing Eq.~(\ref{RGEQEDHO}) and cancels the implicit singlet piece contained
in the term proportional to $\lambda_1$. Not being able to isolate the implicit
piece phenomenologically or to calculate the explicit piece in 
the non-perturbative domain introduces an additional uncertainty, which we now 
argue is rather small.

Perturbation theory provides an order of magnitude estimate if one assumes that
the leading order perturbative coefficient is of typical size and not 
accidentally small. Then one would find for the singlet contribution,
\be
   \delta_{\rm OZI}\sin^2\hat\theta_W(0) \sim \lambda_4 {\hat\alpha\over\pi}
   \left[ 3 {\hat\alpha_s\over\pi} \right]^2 {5\over 324} 
   {11 - 24\zeta(3) \over 33 - 2 n_q} \sim 10^{-6},
\label{errorozi1}
\ee
where the QCD expansion parameter in square brackets has been assumed to have 
grown in the non-perturbative regime to a number of ${\cal O}(1)$, and where 
$n_q = 2$ and $\lambda_4 = 1/20$ correspond to the effective field theory with 
the strange quark integrated out. More generally, based on results of 
Ref.~\cite{vanRitbergen:1998pn} we anticipate that in leading order in $N_C$ 
the singlet terms are of the form,
\be
   \delta_{\rm OZI}\sin^2\hat\theta_W(0) \sim \lambda_4 {\hat\alpha\over\pi}
   \left[ C_A {\hat\alpha_s\over\pi} \right]^n {T_F C_F N_C\over C_A^2} C_n 
   \sim {1\over 90} {\hat\alpha\over\pi} = 2.6 \times 10^{-5},
\label{errorozi2}
\ee
where the QCD group factors are $T_F = 1/2$, $C_F = 4/3$, $C_A = N_C = 3$, and 
where the coefficients $C_n$ are expected to be of ${\cal O}(1)$. 
An alternative form can be written down relative to 
$\Delta\alpha^{(2)}(\bar{m}_s)$ in Eq.~(\ref{ud}),
\be
   \delta_{\rm OZI}\sin^2\hat\theta_W(0) \sim \lambda_4 
   {\Delta\alpha^{(2)}(\bar{m}_s)\over {Q_u^2 + Q_d^2}} {T_F\over C_A} = 
   {3\over 200} \Delta\alpha^{(2)}(\bar{m}_s) = (2.6\mp 1.4) \times 10^{-5},
\label{errorozi3}
\ee
which incidentally gives the same result. These forms exhibits all QED charges 
and leading QCD group factors explicitly, which combined lead to a suppression
of the singlet contribution by two orders of magnitude relative to 
the non-singlet contribution. Thus, the smallness of the estimate in 
Eq.~(\ref{errorozi1}) is in part due to $C_2 \approx - 0.043 \ll 1$ (which may 
or may not reflect the typical size of the other $C_n$), and in part due to 
the suppression factors displayed in Eqs.~(\ref{errorozi2}) and
(\ref{errorozi3}) which will apply at any order. In particular, OZI rule 
violating effects are absent in leading order in the $1/N_C$ 
expansion.

In Appendix~\ref{sec:ozi} we will test these order of magnitude estimates by 
studying the masses and mixings of vector mesons (which strongly dominate 
the real parts of the vector current correlators). The results obtained there 
turn out to be in line with the estimate~(\ref{errorozi1}), but conservatively 
we base our final uncertainty on Eqs.~(\ref{errorozi2}) and (\ref{errorozi3}) 
and take,
\be
   \delta_{\rm OZI}\sin^2\hat\theta_W(0) = \pm 3 \times 10^{-5}.
\label{error3}
\ee

\subsubsection*{Isospin breaking}
So far we have assumed exact isopsin symmetry. Recall that the $SU(3)$ limit 
serves to maximize the RGE running of $\sin^2\theta_W$ by minimizing 
the effective down-type quark masses relative to $\bar{m}_u$. Allowing 
$\bar{m}_d > \bar{m}_u$ can therefore only strengthen this limit. Thus, 
the uncertainty associated with isospin symmetry breaking, 
$\delta_{\rm CVC}\sin^2\hat\theta_W(0)$, is asymmetric.

As for the heavy $\bar{m}_s$ limit, we proceed by considering the hypothetical 
reference case, $\bar{m}_d = \bar{m}_s \neq \bar{m}_u$. In our framework this
corresponds to maximal $SU(2)$ (CVC) violation, {\em i.e.}, $SU(2)$ breaking is
of the same size as $SU(3)$ breaking. The inequality, 
$\Delta\hat\alpha^{(2)}(\bar{m}_s) < 0.00262$ (see Eq.~(\ref{ud})), would be 
replaced by,
\be
   \Delta\hat\alpha^{(1)}(\bar{m}_d) < 0.00262,
\label{u}
\ee
which bounds the up quark contribution for energy scales below the down-type 
quark effective masses. $\lambda_1 = 3/8$ now to be used in 
Eq.~(\ref{solutionHO}) in place of $\lambda_1 = 9/20$ in the isospin symmetric 
case would cause a shift,
\be
   \delta_{\rm CVC}\sin^2\hat\theta_W(0) = 
   - {3\over 40} \Delta\hat\alpha^{(1)}(\bar{m}_d) > - 2 \times 10^{-4}.
\label{errorcvcmax}
\ee
A measure of $SU(2)$ breaking relative to $SU(3)$ breaking is given by 
the ratio,
\be
   \left| {\bar{M}_{K^{*\pm}}^2 - \bar{M}_{K^{*0}}^2 \over 
           \bar{M}_{K^{*\pm}}^2 - \bar{M}_{\rho^0}^2} \right| \approx 0.04,
\ee
which leads to the estimate,
\be
   \delta_{\rm CVC}\sin^2\hat\theta_W(0) = ^{+0}_{-8} \times 10^{-6},
\label{error4}
\ee
and shows that isospin breaking affects our analysis at a very small level.

\subsubsection*{Other uncertainties}
In the perturbative regime we used theory in place of experimental data, which
induces two kinds of uncertainties: purely theoretical ones and parametric ones
from the input quark masses and $\alpha_s$. The former include the errors
associated with the truncation of perturbation theory and with non-perturbative
effects. We estimate their size to about $\pm 7\times 10^{-5}$. However, this
uncertainty is already included in Eq.~(\ref{deltalp}) where it propagates
properly correlated to the error of the low energy weak mixing angle. 
The uncertainties in the quark masses induce an error of $\pm 4\times 10^{-5}$ 
which is dominated by $\hat{m}_c(\hat{m}_c)$. The uncertainty in $\alpha_s$ 
induces an error of the same size. In practice, these parametric uncertainties 
and the one from Eq.~(\ref{deltalp}) are included in global fits to all data 
where these parameters are allowed to float subject to experimental and
theoretical constraints, and where correlations are naturally accounted for. 
The same applies to the experimental uncertainties in 
$\sin^2\hat\theta_W (M_Z)$ and $M_Z$. These have almost no effect on 
$\hat\kappa(0)$, but if the absolute normalization of $\hat{s}^2(0)$ is 
required they induce errors of $\pm 1.4\times 10^{-4}$ and 
$\pm 1.4\times 10^{-5}$, respectively. 

The theoretical uncertainties~(\ref{error1}), (\ref{error2}), (\ref{error3}), 
and (\ref{error4}) added in quadrature yield a total theory error,
\be
   \delta_{\rm theory} \sin^2\hat\theta_W(0) = \pm 7 \times 10^{-5}.
\ee
This is almost an order of magnitude more precise than the result obtained some
time ago in Ref.~\cite{Marciano:1993ep}. Using our results in a global fit to 
precision data yields,
\be
   \sin^2\hat\theta_W(0) = 0.23867 \pm 0.00016.
\label{numvalue}
\ee
The central value coincides with Ref.~\cite{Ferroglia:2003wa} where a seemingly
independent definition of the low energy mixing angle (based on gauge 
invariance and the pinch-technique) is introduced. We will comment more on
the relation between our work and Ref.~\cite{Ferroglia:2003wa} in the following
Section. The uncertainty in Eq.~(\ref{numvalue}) is completely dominated by
the experimental uncertainty,
\be
   \delta \sin^2\hat\theta_W(M_Z) =  \pm 1.5 \times 10^{-4}.
\ee

\section{Other considerations}
\label{sec:finite}
Some time ago, the authors of Refs.~\cite{Czarnecki:1995fw,Czarnecki:2000ic} 
suggested that the use of an appropriate, scale-dependent effective weak mixing angle could 
provide a useful means of comparing the results of various neutral current 
experiments at the $Z^0$-pole and below. By now, it is conventional to compare the value of an effective weak mixing angle extracted from experimental results with its predicted value in the Standard Model (see, {\rm e.g.}, Ref.~\cite{Anthony:2003ub}). More recently, it was 
observed~\cite{Ferroglia:2003wa} that the effective weak mixing angle derived from the sum of $Z$-$\gamma$ mixing diagrams evaluated in the $R_\xi$ gauge and frequently used 
to interpret low energy neutral current experiments is not gauge invariant. 
This $\sin^2\hat\theta(q^2)^{\rm \; eff}$ is defined analogously to 
Eq.~(\ref{eq:sin2eff}) with a scheme- and $q^2$-dependent form factor,
$\hat\kappa(q^2)$,
\be
\label{eq:sin2effq}
   \sin^2\hat\theta(q^2)^{\rm \; eff} \equiv
   {\rm Re} [ \hat{\kappa} (q^2,\mu=M_Z) ] \sin^2\hat\theta_W (\mu = M_Z).
\ee 
In the ${\overline{\rm MS}}$-scheme, both $\hat{\kappa}$ and
$\sin^2\hat\theta_W$ depend on the renormalization scale $\mu$, while
$\hat{\kappa}$ also carries a $q^2$-dependence. The authors of
Ref.~\cite{Ferroglia:2003wa} note that the $\hat{\kappa} (q^2,\mu)$ form factor
naively-defined in terms of $Z$-$\gamma$ mixing depends on 
the choice of electroweak gauge so that the corresponding 
$\sin^2\theta(q^2)^{\rm \; eff}$ is not a physically meaningful quantity. 
By itself, this gauge dependence is not particularly problematic, since for any
physical observable --- such as the parity violating M\o ller asymmetry 
computed in Ref.~\cite{Czarnecki:1995fw} --- it is canceled by the gauge
dependence of other radiative corrections, leaving a gauge independent result. 
Nevertheless, if one wishes to isolate a particular class of radiative 
corrections, such as those entering $\sin^2\hat\theta(q^2)^{\rm \; eff}$, one usually 
prefers to discuss gauge invariant quantities, especially when the comparison 
of different experimental results is involved. 

The authors of Ref.~\cite{Ferroglia:2003wa} show that one may, indeed, obtain 
a gauge independent $\hat{\kappa}(q^2,\mu)$ by including the so-called 
\lq\lq pinch parts" of various one-loop vertex and box diagrams that are 
process independent and that compensate for the gauge dependence of the naive 
$\hat{\kappa}$ form factor. Here, we comment on the relationship between 
$\sin^2\hat{\theta}(q^2)^{\rm \; eff}$ of Ref.~\cite{Ferroglia:2003wa} and
$\sin^2\hat\theta_W (\mu)$ discussed in our work and observe that they are
identical at one-loop order. 

For $|q^2| <  M_W^2$, the gauge invariant form factor $\hat{\kappa}$ of 
Ref.~\cite{Ferroglia:2003wa} is given by,
\bea
\label{eq:sirlina}
   {\hat\kappa}(q^2,\mu)^{\rm PT} &=& 1 + {\alpha\over 2\pi\hat{s}^2_Z}   
   \ln{M_Z^2\over\mu^2} \left[ - {1\over 3}\sum\limits_f N_f^c Q_f 
   (T_f - 2 Q_f \hat{s}^2_Z) + {7\over 2}{\hat c}^2_Z + {1\over 12} \right] \\
   \nonumber && - {\alpha\over 2\pi\hat{s}^2_Z} \left[ 2\sum\limits_f N_f^c Q_f
   (T_f - 2 Q_f \hat{s}^2_Z) I_f(q^2) + \left( {7\over 2}{\hat c}^2_Z +
   {1\over 12}\right) \ln{M_Z^2\over M_W^2} + {{\hat c}^2_Z\over 3} \right],
\eea
where,
\be
   I_f(q^2) = \int_0^1\ dx x (1 - x) \ln{m_f^2 - q^2 x (1 - x)\over M_Z^2},
\ee
and where the PT superscript in Eq.~(\ref{eq:sirlina}) indicates the gauge
invariant ``pinch-technique" definition of the form factor. For $\mu = M_Z$,
the second term on the right hand side of Eq.~(\ref{eq:sirlina}) vanishes.
The integrals $I_f(q^2)$ in the third term generate the large logarithms 
containing fermion masses that one would like to re-sum. As discussed in 
Section~\ref{sec:intro}, this re-summation is accomplished by choosing 
$\mu \sim \sqrt{|q^2|}$ rather than $\mu=M_Z$, thereby eliminating these logarithms from 
${\hat\kappa}(q^2,\mu)^{\rm PT}$ altogether and moving them instead into 
$\sin^2{\hat\theta}_W(\mu)$, which we analyze in this paper. The RGE for $\sin^2{\hat\theta}_W(\mu)$ then provides for the desired re-summation. Similarly, the term proportional to 
$\ln M_Z^2/M_W^2$ in Eq.~(\ref{eq:sirlina}) corresponds to the weak gauge 
sector contributions to the RGE running from $\mu = M_Z$ down to $M_W$. Below 
this scale, the heavy gauge bosons are to be integrated out. 

It is not too surprising that the logarithms appearing, for example, in Eq.~(\ref{kappaLO}), 
are identical to those obtained from the PT since it has been 
shown~\cite{Degrassi:1992ue} that the asymptotic behavior of effective coupling
constants directly constructed from the PT self-energies are automatically 
governed by the renormalization group. Now we observe that even 
the non-logarithmic piece in the third term of Eq.~(\ref{eq:sirlina}) can be
understood in the context of the renormalization group, except that in 
this case it arises from RGE matching rather than RGE running. 
In Ref.~\cite{Ferroglia:2003wa} the ${\hat c}^2_Z/3$ term results from 
combining the pinch parts of the one-loop vertex and box graphs with 
the remaining weak gauge-dependent contributions to the $\gamma Z$-mixing tensor.
The precise value for this $\mu$-independent constant follows from
the requirement that ${\hat\kappa}(q^2,\mu)^{\rm PT}$ be gauge invariant.
In our treatment of the running $\sin^2{\hat\theta}_W(\mu)$, this same constant
is generated by the threshold corrections at $\mu = M_W$ given in
Eqs.~(\ref{eq:thresholda}) and~(\ref{eq:thresholdb}).  Indeed, use of the RGE with appropriate matching conditions may provide a more direct route for obtaining the results of Ref.~\cite{Ferroglia:2003wa} while allowing one to generalize it to include various higher-order effects as we have done.


It follows as a corollary that the PT applied within 
the $\overline{\rm DR}$-scheme (compare the last footnote in 
Section~\ref{sec:matching}) should not yield any constant terms at one-loop 
order. As a particular application, one may consider the correspondence between
the two treatments at $\mu = 0$. Eqs.~(\ref{solutionLO}), 
(\ref{eq:thresholda}), and~(\ref{eq:thresholdb}) show that the relation between
$\sin^2{\hat\theta}_W(0)$ and $ \sin^2{\hat\theta}_W(M_Z)$
\bea
\label{alphanalog}
\sin^2{\hat\theta}_W(0) & =&
   \sin^2{\hat\theta}_W(M_Z) \\
   \nonumber
   &+& {\alpha\over\pi} \left[
   {1\over 6}\sum\limits_f N_f^c Q_f (T_f - 2 Q_f \hat{s}^2_Z)
   \ln{M_Z^2\over m_f^2} - \left( {43\over 24} - {7\over 4} \hat{s}^2_Z \right)
   \ln {M_Z^2\over M_W^2} - {1\over 6} \hat{c}^2_Z \right],
\eea
is the electroweak analog of the relation between the electromagnetic fine structure constant,
$\alpha = \hat\alpha(0)$ and $\hat\alpha(M_Z)$. Note that a similar relation would hold for other definitions of the weak mixing angle and the corresponding definition of the running QED coupling in the same scheme. 
For example, different conventions for the treatment of heavy top quark effects~\cite{Marciano:1990dp} would affect the definitions of $\sin^2{\hat\theta}_W(M_Z)$ and $\hat\alpha(M_Z)$, 
but the right-hand side of relation~(\ref{alphanalog}) would also have to be 
modified with the net effect that the left-hand side would remain unchanged.
Nonetheless, we reiterate that the definition~(\ref{alphanalog}) is gauge invariant because it 
agrees with $\sin^2\hat{\theta}(q^2=0)^{\rm \; eff}$, and that the analysis 
of Section~\ref{sec:rgeHO} has allowed us to incorporate higher order effects 
in $\alpha$ and $\alpha_s$ into $\sin^2\theta_W$. Note also that if 
$\sin^2{\hat\theta}_W(\mu)$ or $\sin^2\hat{\theta}(q^2)^{\rm \; eff}$ are used 
in low $|q^2|$ amplitudes, care must be taken to consistently include other radiative corrections in the same scheme.

While our study has focused on $\sin^2{\hat\theta}_W(\mu)$ appropriate for
low $|q^2|$ processes, it is also worth commenting on the running of the weak 
mixing angle above the weak scale. For $|q^2|\gg M_Z^2$, it is most appropriate
to work in a basis involving the \lq\lq primordial" $SU(2)_L$ and $U(1)_Y$ 
gauge bosons and the $\beta$-functions for $g$ and $g'$. Starting from
Eq.~(\ref{sin2t}) one obtains the RGE,
\be
\label{eq:rgehigh}
   {\hat s}^2 {d{\hat\alpha}\over dt} - {\hat\alpha}{d{\hat s}^2\over dt} =
   {b_2\over\pi} {\hat\alpha}^2 + \sum_j {b_{2j}\over\pi^2}
   {\hat\alpha}^2{\hat\alpha}_j + \cdots,
\ee
where $t = \ln\mu$, and where $b_2$ and $b_{2j}$ are, respectively, the one-
and two-loop $\beta$-function coefficients involving $SU(2)_L$ (see,{\em e.g.},
Ref.~\cite{Jones:1981we}). The solution to Eq.~(\ref{eq:rgehigh}) can be 
written in the same form as Eq.~(\ref{solutionHO}). We note, however, that 
a naive application of the RGE~(\ref{eq:rgehigh}) to scales $\mu \ll M_W$ would
not re-sum all the large logarithms associated with the low $|q^2|$ radiative
corrections. For example, from Eq.~(\ref{eq:rgehigh}) one obtains,
\be
   \lambda_1 = {\sum_q T_q^2\over 2\sum_q Q_q^2}.
\ee
So long as both members of a quark doublet are included in the effective 
theory, this result is equivalent to the expression in Eq.~(\ref{lambda12}), 
since $Q = T + Y$ and ${\rm Tr\,} (TY) = 0$, where $Y$ denotes hypercharge. 
However, for $\mu$ lying between the masses of two doublet members, 
this equivalence no longer holds, and only Eq.~(\ref{lambda12}) will lead to 
a full re-summation of the large logarithms.

\section{Conclusions and outlook}
\label{sec:conclude}
With the completion of the precision electroweak programs at LEP~1, SLC, and
LEP~2, precision measurements of low energy neutral current observables have
taken on added interest in recent years.  A useful way to compare the results from
existing and prospective experiments is to extract the value of the weak 
mixing angle that each would imply, assuming no other physics than that of 
the SM. The extent of their agreement with the SM prediction for this quantity 
provides important information about both the SM as well as the various 
scenarios that might extend it. 

The impact of this low energy precision program depends on both the precision 
of the various experiments as well as that of the SM predictions. 
In this study, we have attempted to refine the latter by giving the appropriate
low energy running weak mixing angle in the ${\overline{\rm MS}}$-scheme, 
$\sin^2\hat\theta_W(\mu)$. By using this quantity and taking $\mu^2\sim |q^2|$,
one is able to re-sum various logarithmically-enhanced contributions that would
otherwise appear in the radiative corrections for $|q^2| \ll M_W^2$, thereby
reducing the truncation error associated with the perturbative expansion.
At one-loop order, this re-summation reproduces the result of
Ref.~\cite{Ferroglia:2003wa}, but we have been able to generalize that work 
to include higher order contributions in $\alpha$ and $\alpha_s$. We have also 
provided an extensive analysis of the non-perturbative hadronic contributions 
to $\sin^2\hat\theta_W (\mu)$ for $\mu\sim 0$ and argued that the associated 
uncertainties enter below the $10^{-4}$ level. 

\begin{figure}[t]
\psfig{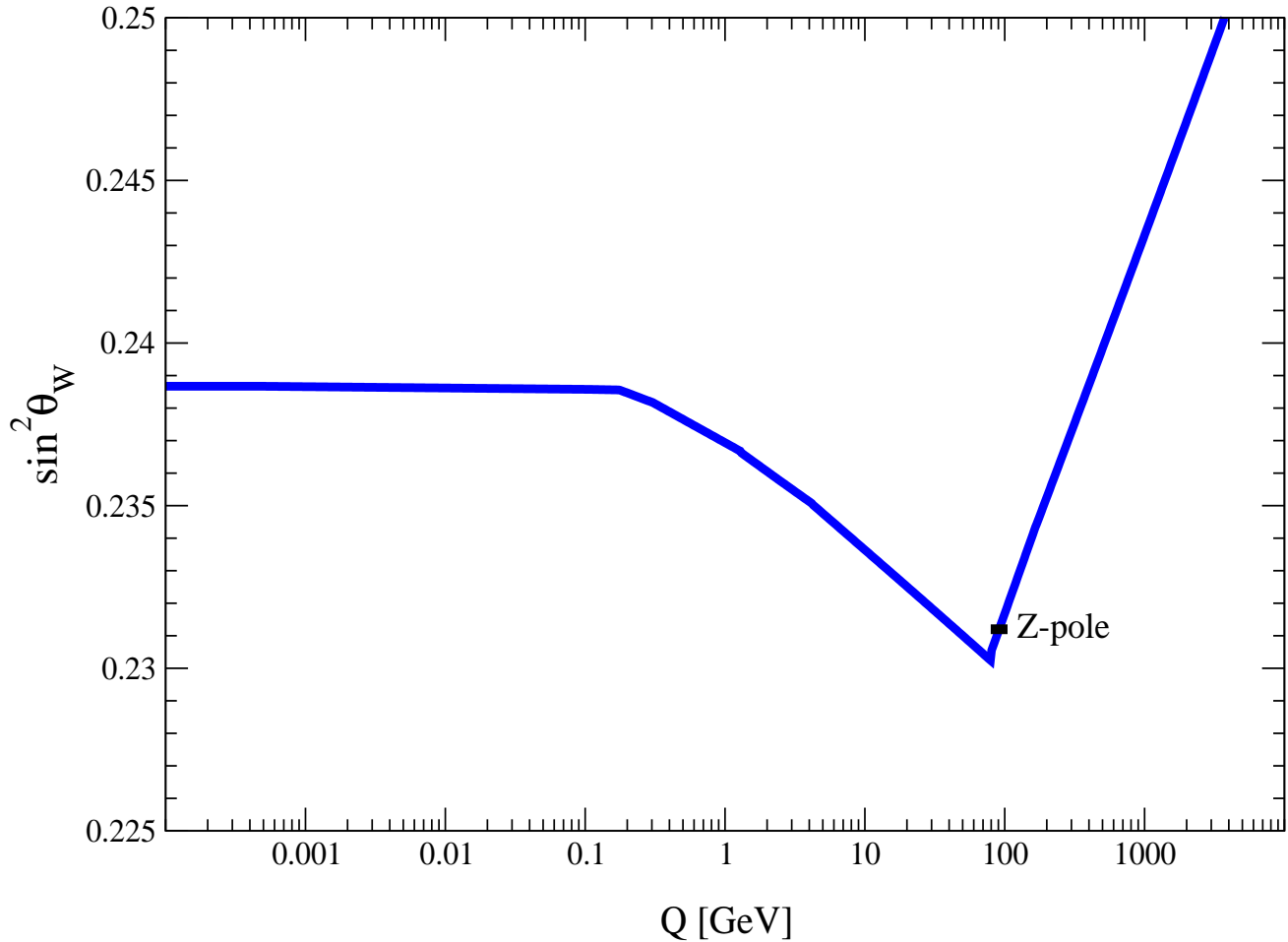}
\caption{Scale dependence of the weak mixing angle in the 
\msbar~renormalization scheme.}
\label{fig}
\end{figure}

The resulting scale-dependence of $\sin^2\hat\theta_W (\mu)$ for 
$\mu=\sqrt{|q^2|}$ with $q^2$ being the four-momentum transfer squared is 
shown in Fig.~\ref{fig}. The various discontinuities in the curve correspond 
to the thresholds discussed above, while the size of the theoretical 
uncertainty in the curve corresponds to its thickness. 

As a particular application, we 
obtained a definition of the mixing angle in the Thomson limit, 
$\sin^2\theta_W \equiv \sin^2\hat\theta_W(0)$ whose relation with the value determined at the $Z$-pole
$\sin^2\hat\theta_W(M_Z)$ is the electroweak analog of the relation between the fine structure $\alpha =\hat\alpha(0)$ constant and $\hat\alpha(M_Z)$. This definition also
coincides with the gauge invariant definition recently constructed in 
Ref.~\cite{Ferroglia:2003wa}, and its numerical value is
\be
   \sin^2\theta_W = 0.23867 \pm 0.00016,
\ee
where the error is dominated by the experimental error from $Z$ pole
measurements. From its relation to the $\overline{\rm MS}$-scheme mixing angle at 
the $Z$ scale, $\sin^2\hat\theta_W(M_Z)$, by Eq.~(\ref{sin2def}), we obtain
\be
   \Delta\hat\kappa(0) = 0.03232 \pm 0.00029,
\ee
where now the error is purely theoretical. Finally, using 
the relation~\cite{Gambino:1993dd} between $\sin^2\hat\theta_W(M_Z)$ and
the effective leptonic mixing angle, $\sin^2\theta_f^{\rm \; eff}$, defined in 
Eq.~(\ref{eq:sin2eff}), we obtain,
\be
   \sin^2\theta_W - \sin^2\theta_\ell^{\rm \; eff} = 0.00718 \pm 0.00007.
\ee
The error in this relation (which is the main result of this work) is an order 
of magnitude below current and anticipated experimental uncertainties and considerably smaller than the uncertainty quoted in Refs.~\cite{Marciano:1993ep}. 

To illustrate the impact of this result on particular observables, we consider the weak charge of the proton, $Q_W(p)$, that will be determined using parity-violating elastic $ep$ scattering at the Jefferson Lab. Up-dating the recent analysis of Ref.~\cite{Erler:2003yk}, we obtain the Standard Model prediction
\bea
\nonumber
Q_W(p) &=& 0.0713 \pm 0.0006~({\rm input}) \pm 0.0003~(\Delta\hat\kappa) \pm 0.0005~(Z\gamma\ {\rm box}) \pm 0.0001~(WW\ {\rm box}) \\
\label{eq:qwp}
&=& 0.0713\pm 0.0008,
\eea
where the experimental uncertainties in $\sin^2\hat\theta_W(M_Z)$ and $m_t$~(\lq\lq input"), the theoretical hadronic uncertainties in $\Delta\hat\kappa(0)$ and the $Z\gamma$ box graph, and the uncertainty from unknown higher-order perturbative QCD contributions to the $WW$ box graphs are shown separately and combined in quadrature in the second line. Use of the previous estimate for the uncertainty in $\Delta\hat\kappa(0)$ of $\pm 0.0025$ would lead to a theoretical (total) error in the $Q_W(p)$ prediction roughly five (three) times larger than in Eq.~(\ref{eq:qwp}), and in this case, one could neglect the uncertainties associated with other radiative corrections. In light of our analysis, however, the uncertainty in the Standard Model prediction is now three and a half times smaller than the anticipated experimental error, and theoretical uncertainties associated with hadronic contributions to other radiative corrections become relatively more important.  

In the same vein,  the interpretation of the prospective parity violating deep inelastic measurements will require a careful analysis of higher twist and isospin breaking corrections, especially given that the latter may be responsible for the present discrepancy between the NuTeV result for $\sin^2\hat\theta_W (\mu\sim 3\ {\rm GeV})$ and the SM prediction. Our analysis applies to the deep inelastic regime, as well,
where due to the higher energies involved, the uncertainty in $\sin^2\hat\theta_W (\mu)$ is even much smaller since no complications from non-perturbative contributions arise in this case. Given the level of experimental effort required to carry out these precise low energy measurements, performing a theoretical analysis of these effects at the level we have attempted to do here for the weak mixing angle seems well-worth the effort.

\section*{Acknowledgments:}
We are happy to thank Krishna Kumar, Paul Langacker, Philip Mannheim, Robert 
McKeown, Michael Pennington, Alberto Sirlin, and Peter Zerwas for very helpful 
discussions, and Peter Zerwas also for a careful reading of the manuscript. 
We are grateful for the support from the Institute for Nuclear Theory at 
the University of Washington, where this work was initiated. MR-M thanks 
the Physics Institute at UNAM for support and hospitality during a visit 
to undertake this project. This work was supported in part by U.S.\ Department 
of Energy contracts DE--FG02--00ER41146, DE--FG03--02ER41215, and 
DE--FG03--00ER41132, by the National Science Foundation Award PHY00--71856, by 
CONACYT (M\'exico) contract 42026--F, and by  DGAPA--UNAM contract PAPIIT 
IN112902.

\appendix
\section{Updated value for the mixing angle $\epsilon$}
\label{sec:epsilon} 
Here we determine the phenomenological value of the $\omega$-$\phi$ mixing 
angle, $\epsilon$, defined by,
\be
\label{epsilon}
 |\omega\rangle = \cos\epsilon {|\bar{u}u\rangle+|\bar{d}d\rangle\over\sqrt{2}}
 - \sin\epsilon |\bar{s}s\rangle, \hspace{50pt}
 |\phi\rangle   = \sin\epsilon {|\bar{u}u\rangle+|\bar{d}d\rangle\over\sqrt{2}}
 + \cos\epsilon |\bar{s}s\rangle,
\ee
where $\epsilon = 0$ is referred to as ideal mixing. One way to extract it is 
to use $SU(3)$ flavor symmetry and first order breaking applied to the vector 
meson octet mass spectrum. An advantage of this method is that it can be 
calibrated against the ground state baryon octet, for which Fermi statistics 
precludes the mixing with an $SU(3)$ singlet state. The mass of the $SU(2)$ 
singlet, $M_\Lambda$, is therefore predicted in terms of the masses of 
the other electrically neutral octet members,
\be
   M_\Lambda = {1\over 3} (2 M_n + 2 M_{\Xi^0} - M_{\Sigma^0}) = 
   1105.4\mbox{ MeV},
\ee
which reproduces the experimental value~\cite{Eidelman:2004wy},
$M_\Lambda = 1115.7$~MeV, within 0.93\%. Analogously, the Gell-Mann-Okubo mass 
formula~\cite{Okubo:1961jc,Gell-Mann:1962xb} yields the octet-octet component 
of the mass matrix for the isosinglet ground state vector mesons,
\be
   \bar{M}_{88}^2 = {1\over 3} (4 \bar{M}^2_{K^{*0}} - \bar{M}^2_{\rho^0}) = 
   (933.69\mbox{ MeV})^2 \times [1\pm 0.0008 \pm 0.0020 \pm 0.0121 \pm 0.0093].
\label{m88}
\ee
The first error is from the experimental uncertainty in the masses, which are 
taken from Ref.~\cite{Eidelman:2004wy} except for the mass, 
$M_{\rho^0} = 775.74 \pm 0.65$~MeV, and the width, 
$\Gamma_{\rho^0} = 145.3 \pm 1.4$~MeV, of the $\rho^0$ resonance for which we 
averaged Refs.~\cite{Barkov:1985ac,Akhmetshin:2003zn}. The broadness of some of
the resonances involved introduces an ambiguity as for what definition of mass 
one should use in the Gell-Mann-Okubo formula. For the central value we have 
chosen the peak position,
\be
   \bar{M} = {M\over\sqrt[4]{1 + {\Gamma^2\over M^2}}},
\ee
where $M$ and $\Gamma$ correspond to the usual definition of a relativistic
Breit-Wigner resonance form with an $s$-dependent width, {\em i.e.} the one
used by the Particle Data Group~\cite{Eidelman:2004wy}. The second error in 
Eq.~(\ref{m88}) reflects the size of the shift obtained by replacing $\bar{M}$ 
by $M$. The third error is due to possible isospin breaking effects which we 
estimated by using $\bar{M}^2_{K^{*\pm}}$ in place of $\bar{M}^2_{K^{*0}}$.  
The last error quantifies the limitation of the method and is given by 
the calibration against the baryon octet as discussed above. Adding all errors 
in quadrature, we obtain for the $SU(3)$ octet-singlet mixing 
angle~\cite{Eidelman:2004wy}, $\theta_V$,
\be
   \tan^2\theta_V = \frac{\bar{M}_{88}^2     - \bar{M}_{\phi}^2}
                         {\bar{M}_{\omega}^2 - \bar{M}_{88}^2} 
                  = 0.646^{-0.081}_{+0.090},
\ee
which translates into the two solutions\footnote{Since $\Gamma_\phi \ll M_\phi$
and $\Gamma_\omega \ll M_\omega$, the complex phase in $\epsilon$ can safely be
neglected.}, $\epsilon_1 = 0.061\mp 0.032$ and $\epsilon_2 = - 1.292\pm 0.032$.
Alternatively, one can compare the branching ratios of the $\omega$ and $\phi$
resonances decaying into $\pi^0 \gamma$~\cite{Jain:1987sz},
\be
   \tan^2\epsilon = {M_\phi^3\over M_\omega^3} 
   \left( \frac{M_\omega^2 - M_{\pi^0}^2}{M_\phi^2 - M_{\pi^0}^2} \right)^3 
   \frac{{\cal B}(\phi \to \pi^0 \gamma)}{{\cal B}(\omega \to \pi^0 \gamma)} 
   {\Gamma_\phi\over\Gamma_\omega} = (3.01 \pm 0.26) \times 10^{-3},
\ee 
which gives, $|\epsilon| = 0.0548 \pm 0.0024$.  Comparison with the previous 
method singles out the solution,
\be
   \epsilon = + 0.0548 \pm 0.0024.
\ee
The sign and magnitude are also consistent with various other 
methods~\cite{Arafune:1977pk} and the previous analysis in 
Ref.~\cite{Jain:1987sz}.

\section{Phenomenological approach to OZI suppression}
\label{sec:ozi} 
In Section~\ref{sec:uncertainties} we argued that the OZI rule can at least 
partly be understood as a result of group theoretical suppression factors 
relative to OZI allowed processes. Using the result of 
Appendix~\ref{sec:epsilon}, we now wish to study OZI rule violating 
contributions to the mass matrix of ground state vector mesons. These mesons 
dominate the electromagnetic current correlator at hadronic scales, and should 
therefore serve as a means to quantify OZI rule suppressions 
phenomenologically. Throughout this Appendix we work in the isospin symmetric 
limit.

In the flavor basis, $(|\bar{u}u\rangle,|\bar{d}d\rangle,|\bar{s}s\rangle)$,
we write the mass matrix in a form which is similar to the one discussed in
Ref.~\cite{DeRujula:1975ge},
\be
   M = \left( \ba{ccc} A + B &   B   &   B + C   \\ 
                         B   & A + B &   B + C   \\ 
                       B + C & B + C & A + B + D \ea \right).
\ee
The parameters $A$ and $B$ respect $SU(3)$ symmetry, while $C$ and $D$ break 
it. The off-diagonal elements, $B$ and $C$, parametrize flavor transitions, and
can only be generated by QCD annihilation (singlet) diagrams. The parameter $D$
receives contributions from the strange quark mass, as well as dynamical 
contributions of both singlet and non-singlet type. The difference to 
Ref.~\cite{DeRujula:1975ge} is that there $C = 0$, and $B = B(\mu)$ is a scale 
dependent singlet function, while we {\em define\/} all entries of $M$ as 
constants without specifying their relation to the scale dependent current 
correlators. In the isospin basis, $(|(\bar{u}u-\bar{d}d)/\sqrt{2}\rangle, 
|(\bar{u}u+\bar{d}d)/\sqrt{2}\rangle, |\bar{s}s\rangle)$, $M$ reads,
\be
   M = \left( \ba{ccc} A &       0          &       0          \\ 
                       0 &   A + 2 B        & \sqrt{2} (B + C) \\ 
                       0 & \sqrt{2} (B + C) &   A + B + D      \ea \right),
\ee
so that we can identify, $A = \bar{M}^2_{\rho^0}$. The trace of $M$ then yields
the condition, 
\be
   3 B + D = \bar{M}^2_\omega + \bar{M}^2_\phi - 2 \bar{M}^2_{\rho^0},
\ee
and $\epsilon$ obtained in Appendix~\ref{sec:epsilon} gives the constraint,
\be
   \tan 2 \epsilon = \sqrt{8} {B + C \over D - B}.
\ee
The final relation,
\be
   (\bar{M}^2_\phi - \bar{M}^2_\omega)^2 = 
   \left[ 3 (B + C) - {1\over 3} (C + D) \right]^2 + {8\over 9} (C + D)^2,
\ee
shows that in the $SU(3)$ limit, $C = D = 0$, singlet diagrams associated with
$B$ would split $M^2_\phi$ from $M^2_{\rho^0} = M^2_\omega$ in much the same 
way as triangle anomaly diagrams would split $M^2_{\eta^\prime}$ from 
$M^2_\eta = M^2_\pi$. Taking into account that $\epsilon\ll 1$, we can 
approximate,
\be
   B \approx {\bar{M}^2_\omega - \bar{M}^2_{\rho^0} \over 2}, \hspace{100pt}
   D \approx \bar{M}^2_\phi - {\bar{M}^2_\omega + \bar{M}^2_{\rho^0} \over 2},
\ee
which is correct up to ${\cal O}(\epsilon^2)$. Thus, in the limit of ideal
mixing, $B$ drives the splitting of $M^2_\omega$ from $M^2_{\rho^0}$ instead.
In any case, the singlet contribution associated to $B$ is very small compared 
to $A$ and $D$. More relevant for this work is the $SU(3)$ breaking singlet 
parameter,
\be
   C \approx {\bar{M}^2_\phi - \bar{M}^2_\omega \over \sqrt{8}} \tan 2\epsilon
           - {\bar{M}^2_\omega - \bar{M}^2_{\rho^0} \over 2},
\ee
which reduces to $C = - B$ in the ideal mixing case, $\epsilon = 0$. 
Numerically, we have,
\be\ba{ll}
   A =     (769\mbox{ MeV})^2, & B = (105\mbox{ MeV})^2, \\
   C = (\;\; 74\mbox{ MeV})^2, & D = (660\mbox{ MeV})^2. 
\ea\ee
There is a second solution in which $|B|$, $|C|$, and $|D|$ are all comparable 
and where $B \approx - C$ to ensure $\epsilon \ll 1$. It also has $D < 0$, 
although the strange quark mass is expected to give the dominant (positive) 
contribution. Therefore we discard this solution. We can also roughly estimate 
the singlet component contained in $D$ by relating it to $M_{K^{*0}}$,
\be
   D_{\rm singlet} \sim D - 2 (\bar{M}^2_{K^{*0}} - \bar{M}^2_{\rho^0}) 
   = (123\mbox{ MeV})^2,
\ee
which is of similar size as $B$ and $C$. Thus, singlet contributions to vector
meson masses are generally suppressed by more than an order of magnitude beyond
the QCD suppression factors discussed in Section~\ref{sec:uncertainties} 
indicating that the smallness of the known coefficient, $C_2 \ll 1$, may indeed
be a generic feature that persists at higher orders. Qualitatively the same 
results are obtained if linear mass relations are used in place of masses 
squared, except that singlet contributions appear even more suppressed.

\end{document}